\documentclass[notitlepage,floats,aps,nofootinbib,preprintnumbers,superscriptaddress,twocolumn,prl,10pt,balancelastpage]{revtex4-1}

\usepackage{amsmath,amssymb,amsfonts}
\usepackage{hyperref}
\hypersetup{colorlinks,citecolor=blue,urlcolor=blue,linkcolor=blue}
\usepackage{graphicx}
\usepackage{xcolor}
\usepackage{subfigure}
\usepackage{nicefrac}
\usepackage{mathrsfs}
\usepackage{mdframed}
\usepackage{ulem}
\usepackage{units}

\def\beq{\begin{equation}\begin{aligned}}
\def\eeq{\end{aligned}\end{equation}}

\begin{document}

\preprint{DESY-17-210}

\title{Protecting the Axion with Local Baryon Number}

\author{Michael Duerr} 
\author{Kai Schmidt-Hoberg}
\affiliation{Deutsches Elektronen-Synchrotron DESY, Notkestrasse 85, 22607 Hamburg, Germany}
\author{James Unwin}
\affiliation{Deutsches Elektronen-Synchrotron DESY, Notkestrasse 85, 22607 Hamburg, Germany}
\affiliation{Department of Physics, University of Illinois at Chicago, Chicago, IL 60607, USA}

\begin{abstract} 
The Peccei--Quinn~(PQ) solution to the Strong CP Problem is expected to fail unless the global symmetry U(1)${}_{\rm PQ}$ is protected from Planck-scale operators up to high mass dimension. Suitable protection can be achieved if the PQ symmetry is an automatic consequence of some gauge symmetry. We highlight that if baryon number is promoted to a gauge symmetry, the exotic fermions needed for anomaly cancellation can elegantly provide an implementation of the Kim--Shifman--Vainshtein--Zakharov  `hidden axion' mechanism with a PQ symmetry protected from Planck-scale physics.
\end{abstract}

\maketitle


\section{Introduction}
\label{sec:Introduction}
In principle one expects that the QCD Lagrangian should contain a CP-violating term of the form
\beq
\mathcal{L}_{\rm QCD}\supset -\bar{\theta}~\frac{\alpha_S}{8\pi}G_{\mu\nu}\widetilde{G}^{\mu\nu}\,.
\eeq
This would imply a non-vanishing neutron electric-dipole moment $d_n\simeq5.2\times10^{-16}\bar{\theta}~e$~cm~\citep{Crewther:1979pi}, which has been constrained to be $d_n<\unit[3.0\times10^{-26} e]{cm}$ at 90 \% C.L.~by experimental searches \citep{Afach:2015sja}. Satisfying this experimental upper limit  on $d_n$ requires  $\bar{\theta} \lesssim\bar{\theta}_{\rm lim}\simeq10^{-10}$ and the unexplained smallness of $\bar{\theta}$ is the Strong CP Problem.

Fortunately, there is an elegant resolution due to Peccei and Quinn (PQ) \cite{Peccei:1977hh,Peccei:1977ur} in which $\bar{\theta}$ is promoted to a dynamical field and the scalar potential is minimised for $\bar{\theta}=0$.  The PQ mechanism involves a global symmetry U(1)${}_{\rm PQ}$ which is spontaneously broken by a vacuum expectation value (VEV) and explicitly broken by the QCD chiral anomaly. If in addition U(1)${}_{\rm PQ}$ is explicitly broken by other sources, then generically $\bar\theta\neq0$. One potential concern in this context is that continuous global symmetries are expected to be explicitly broken by Planck-scale ($M_{\rm Pl}$) physics~\citep[and references therein]{Banks:2010zn}. Indeed, unless Planck-scale suppressed operators with mass dimension $D\lesssim9$ are absent, then generically $\bar\theta>\bar\theta_{\rm lim}$, in conflict with observation~\citep{Kamionkowski:1992mf,Holman:1992us}. Thus for the  PQ mechanism to be successful these PQ violating operators must be forbidden or highly suppressed. Manners of forbidding these PQ violating Planck-scale operators include imposing discrete symmetries, or embedding U(1)${}_{\rm PQ}$ into a local symmetry or GUT~\citep{Georgi:1981pu,Holman:1992us,Barr:1992qq,Randall:1992ut,Redi:2016esr,DiLuzio:2017tjx}.   

Here we consider models with the gauge symmetry  
\begin{equation}
{\rm SU(3)}\times{\rm SU(2)}_L\times{\rm U(1)}_Y\times{\rm U(1)}_{\mathscr{B}}\,,
\label{gauge}
\end{equation}
where U(1)${}_{\mathscr{B}}$  is {\it generalised baryon number}, under which the Standard Model quarks carry charge $\nicefrac{1}{3}$, whilst other Standard Model fields are neutral. Promoting the accidental U(1) symmetries of the Standard Model to gauge symmetries is an interesting idea with roots in the work of~\cite{Lee:1955vk,Pais:1973mi,Foot:1989ts}. Recently such models have received a marked growth in interest following the explicit models of Fileviez Perez and Wise \cite{FileviezPerez:2010gw,FileviezPerez:2011pt} (see also~\citep{Duerr:2013dza}).
The quantum numbers of the Standard Model fermions (with three right-handed neutrinos $N_R$) are 
\beq
Q_L&=(\mathbf{3},\mathbf{2},\nicefrac{1}{3},\nicefrac{1}{3})_{\times3}\,,    &L_L&=(\mathbf{1},\mathbf{2},-1,0)_{\times3}\,,\\
u_R&=(\mathbf{3},\mathbf{1},\nicefrac{4}{3},\nicefrac{1}{3})_{\times3}\,,   &e_R&=(\mathbf{1},\mathbf{1},-2,0)_{\times3}\,,\\
d_R&=(\mathbf{3},\mathbf{1},\nicefrac{-2}{3},\nicefrac{1}{3})_{\times3}\,,  &N_R&=(\mathbf{1},\mathbf{1},0,0)_{\times3}\,, 
\eeq
with notation $(\mathbf{d_3},\mathbf{d_2},q_y,q_{\mathscr{B}})_{\times n}$ where $\mathbf{d_N}$ is the dimension of the representation under SU($N$), the $q_i$ are U(1)${}_i$ charges and the subscript $\times3$ indicates the number of families. The quantum numbers of the Higgs are $H=(\mathbf{1},\mathbf{2},1,0)_{\times 1}$. In the Standard Model global baryon number U(1)${}_B$ is anomalous with non-vanishing anomaly coefficients
 \beq
\mathcal{A}_{{\rm U(1)}_B\times{\rm SU}(2)^2} &=\nicefrac{3}{2}\,, \qquad
\mathcal{A}_{{\rm U(1)}_B\times{\rm U}(1)_Y^2}&=-6\,.
\label{SMa}
\eeq
An anomaly-free theory of gauged baryon number therefore requires new chiral fermions transforming under the Standard Model gauge group as well as U(1)${}_\mathscr{B}$.

In this work we highlight that the PQ symmetry can be protected from $M_{\rm Pl}$ operators by U(1)${}_{\mathscr{B}}$ gauge invariance.\footnote{Associating a PQ symmetry to local baryon number was briefly remarked on in an early paper of Foot, Joshi, and Lew \cite{Foot:1989ts}.} This is a well motivated example of the case that U(1)${}_{\rm PQ}$ arises as an accidental symmetry due to a gauge symmetry. Moreover, the chiral fermions needed for anomaly cancellation can naturally yield the field content to implement the Kim--Shifman--Vainshtein--Zakharov  (KSVZ) mechanism for `hiding' the axion~\citep{Kim:1979if,Shifman:1979if}.  Since baryon number is intimately tied to the QCD sector, it is aesthetically pleasing to connect the resolution of the Strong CP Problem to a baryonic symmetry. 

\section{Gauge-embedded PQ Symmetries}
\label{sec:automaticPQsymmetries}
The PQ mechanism dynamically sets $\bar\theta\approx0$ provided there is some global U(1)${}_{\rm PQ} $ for which the main source of explicit breaking is due to the QCD chiral anomaly, $\mathcal{A}_{{\rm U(1)}_{\rm PQ}\times{\rm SU}(3)^2} \neq0$. In the following we discuss how such an anomalous global U(1) can arise in models with gauged baryon number U(1)${}_{\mathscr{B}}$.

We supplement the Standard Model with exotic fermions charged under the Standard Model gauge group as well as U(1)${}_{\mathscr{B}}$. If cross-terms involving both Standard Model quarks and exotics are forbidden by gauge invariance and the choice of particle content, the model exhibits a global `exotics' symmetry U(1)${}_X$ which is independent of global Standard Model baryon number U(1)${}_B$. Nevertheless, the baryon and exotics sectors are linked by the underlying U(1)${}_{\mathscr{B}}$ symmetry. It is insightful to fix the global U(1)${}_X$ and U(1)${}_B$ charges to match the corresponding U(1)${}_{\mathscr{B}}$ charges. Given this choice, the cancellation of the U(1)${}_{\mathscr{B}}\times \mathcal{G}^2$ gauge anomaly implies that the anomalies of the global symmetries with some gauge group $\mathcal{G}$ must be equal and opposite:
\beq
\mathcal{A}_{{\rm U(1)}_{\mathscr{B}}\times\mathcal{G}^2}=0 ~~~\Rightarrow ~~~
\mathcal{A}_{{\rm U(1)}_B\times\mathcal{G}^2}=-\mathcal{A}_{{\rm U(1)}_X\times\mathcal{G}^2}.
\label{condX}
\eeq
With only the Standard Model quarks carrying global baryon number, U(1)${}_B$ does not have a QCD anomaly ($\mathcal{A}_{{\rm U(1)}_B\times{\rm SU}(3)^2} =0$) and hence also  $\mathcal{A}_{{\rm U(1)}_X\times{\rm SU}(3)^2}=0$,  impeding an identification of exotics number U(1)${}_X$ with the PQ symmetry.
 
One can however easily envision two independent global `exotics' symmetries, which we call suggestively U(1)${}_{\rm PQ}$ and  U(1)$'{}_{\rm PQ}$. Independent global symmetries occur if there are no cross-terms between sets of fermions. 
With the  U(1)${}_{\mathscr{B}}$ induced normalisation the anomalies of these two individual global symmetries are related to the anomaly of global exotics number by
\beq
\mathcal{A}_{{\rm U(1)}_X\times\mathcal{G}^2}=
\mathcal{A}_{{\rm U(1)}_{\rm PQ}\times\mathcal{G}^2}+
\mathcal{A}_{{\rm U(1)}'_{\rm PQ}\times\mathcal{G}^2}.
\eeq
By eq.~(\ref{condX}) it remains that $\mathcal{A}_{{\rm U(1)}_X\times\mathcal{G}^2}=0$, however with appropriate charge assignments one can arrange for 
\beq
\mathcal{A}_{{\rm U(1)}_{\rm PQ}\times {\rm SU(3)}^2}=-
\mathcal{A}_{{\rm U(1)}'_{\rm PQ}\times  {\rm SU(3)}^2}\neq0\,,
\eeq
and thus these global symmetries are suitable for implementing the PQ mechanism.

Spontaneous breaking of the two independent global PQ symmetries will lead to two Nambu--Goldstone bosons. The physical axion $a$ will be an admixture of these two Goldstone bosons while the orthogonal field will be eaten by the U(1)${}_{\mathscr{B}}$ gauge boson $Z'$ resulting in a non-zero mass $m_{Z'}$. The effective axion decay constant $f_a$ is given by~\citep{Fukuda:2017ylt}
\beq
f_a=\frac{ff'}{\sqrt{f^2+f'{}^2}} \,, 
\eeq 
where $f$ and $f'$ are the breaking scales of U(1)${}_{\rm PQ}$ and U(1)${}_{\rm PQ}^{'}$.
The `axion window' in which $f_a$ is appropriate to avoid cosmological and astrophysical limits is roughly $10^{9}~{\rm GeV}\lesssim f_a\lesssim10^{12}$ GeV~\citep{Patrignani:2016xqp}. The axion obtains a mass from the QCD chiral anomaly and a contribution  $\Delta_a$ from Planck-scale breaking  
  \beq
m_a^2\simeq (m_a^\text{QCD})^2 +\Delta_a ^2\,.
\label{ma}
\eeq
For the case that the leading $M_{\rm Pl}$ breaking operator is mass dimension  $D$, the correction $\Delta_a$ is of order~\citep{Kamionkowski:1992mf,Barr:1992qq}
\beq
\Delta_a^2  \sim g\cos\delta M_{\rm Pl}^2\left(\frac{f_a}{M_{\rm Pl}}\right)^{D-2},
\label{Delta}\eeq
where $g$ is a coupling associated to gravitational interactions and $\delta$ is the rotation of the phase of the fermion mass matrix into $\bar\theta$. 
Breaking of U(1)${}_{\rm PQ}$ by gravitational effects leads to  $\bar\theta\propto\Delta_a^2$ which is generically non-zero. 
Specifically, (for $\Delta_a^2 \ll m_a^2$) the Planck-scale breaking of the PQ symmetry leads to
\beq
\bar\theta\simeq \frac{\Delta_a^2}{m_a^2} \tan\delta \sim g\sin\delta \frac{ M_{\rm Pl}^2}{m_a^2}\left(\frac{f_a}{M_{\rm Pl}}\right)^{D-2} \,.
\eeq
For $g,\delta \sim \mathcal{O}(1)$ and $f_a \sim 10^{11}$~GeV the dimension $D$ of the gravity induced operator should be $D \gtrsim 10$ in order to ensure 
$\bar\theta \leq \bar\theta_\text{lim}$.
We note in passing that there are two potential caveats: further \textit{ad-hoc} symmetries could forbid these operators up to some order or, alternatively, if $g\ll1$ Planck-scale breaking may not be important. There are some arguments in the literature that these operators could potentially be exponentially suppressed~\citep{Kallosh:1995hi}, however, it is far from clear that this is the case and na\"ively one might expect gravitationally induced operators with $\mathcal{O}(1)$ coefficients. Thus, in general protection from Planck-scale symmetry violation should be considered an important requirement for successful implementations of the PQ mechanism.

\section{PQ Protection from  \texorpdfstring{$\mathbf{U(1)_{\mathscr{B}}}$}{U1B}}
\label{sec:protectingPQsymmetries}
As was discussed in the last section, to implement a protected PQ symmetry with gauged baryon number one needs at least two pairs of fermion exotics that can be associated with two accidental PQ symmetries. Moreover, one requires that at least one exotic fermion pair transforms non-trivially under SU(2)${}_L$ in order to cancel the SU(2)${}^2_L\times$U(1)${}_{\mathscr{B}}$ anomaly. However, if the fermion exotics carry SU(3) colour and are chiral under U(1)${}_{\mathscr{B}}$ one cannot simultaneously cancel the SU(2)${}_L^2\times$U(1)${}_{\mathscr{B}}$ and SU(3)${}^2\times$U(1)${}_{\mathscr{B}}$ anomalies if all the exotic fermions transform in the same SU(2)${}_L$ representation. To allow simple mass terms for the fermion exotics, we limit ourselves to pairs that are vector-like under the Standard Model gauge group.

The Planck-scale operators which dominantly cause $\bar\theta$ to deviate from zero are those involving only scalars with high scale VEVs, while operators involving fields that do not obtain VEVs are expected to be subdominant~\citep{Dobrescu:1996jp}.  Therefore for a given PQ model it is important to determine the leading scalar operator which violates the PQ symmetry. 

Consider the following set of $n$ fermion pairs,\footnote{A new family of quarks which are vector-like under the Standard Model gauge group, as studied in \citep{FileviezPerez:2011pt}, corresponds to $n=3$ pairs with $\boldsymbol{\lambda_1} = \mathbf{2}$ and $\boldsymbol{\lambda_2} = \boldsymbol{\lambda_3} = \mathbf{1}$.}
 which we label by the dimension $\lambda_i\in\mathbb{N}$ of their SU(2)${}_L$ representation,
\begin{align}\label{eq:generalSolution}
\lambda_1{}_L&=(\mathbf{3},\boldsymbol{\lambda_1},Y_1,B_1), &\lambda_1{}_R &=(\mathbf{3},\boldsymbol{\lambda_1},Y_1,B_1^\prime),\nonumber \\ 
& \vdots & & \vdots \\
\lambda_n{}_L&=(\mathbf{3},\boldsymbol{\lambda_n},Y_n,B_n), &\lambda_n{}_R &=(\mathbf{3},\boldsymbol{\lambda_n},Y_n,B_n^\prime)\nonumber,
\end{align}
with $Y_i,B_i,B_i' \in\mathbb{Q}$. Note that whilst no theorem precludes irrational charges in gauge theories, it has been argued that irrational charges are inconsistent with theories of quantum gravity~\citep{Banks:2010zn}. The simultaneous cancellation of the SU(2)${}^2\times$U(1)${}_{\mathscr{B}}$ as well as the SU(3)${}^2\times$U(1)${}_{\mathscr{B}}$ anomalies enforces the relation 
\begin{equation}
 \sum_{i=1}^n \lambda_i \left( B_i - B_i^\prime \right)  =0.
\end{equation}
Other anomalies further constrain possible charges of the fermion exotics. 

In this setup one generally needs up to $n$ new scalars $\Phi_i$ to give mass to all $n$ pairs of fermion exotics. Diagonal mass terms can be obtained with  $\Phi_i = (\mathbf{1},\mathbf{1},0,B_i - B_i^\prime)$. To avoid the presence of massless Goldstone bosons, we limit ourselves to cases in which two exotic scalars can give mass to all fermion exotics. This can be achieved if some of the differences $\Delta_i \equiv B_i - B_i^\prime$ are chosen to be equal, resulting in two scalars with U(1)${}_{\mathscr{B}}$ charges $\Delta_1$ and $\Delta_2$. Then, one of the Goldstone bosons is eaten by the U(1)${}_\mathscr{B}$ gauge boson, and the other is the axion.

In this scenario the lowest-order scalar operator which breaks the anomalous combination of the global PQ symmetries (but respects the gauged combination) occurs at mass dimension $D$ with
\begin{equation}
 D \le \sum_{i=1}^n  \lambda_i \,.
\end{equation}
This inequality is saturated if the following are coprime:
\beq
\sum_{B_i - B_i^\prime = \Delta_1} \lambda_i \quad {\rm and}\quad \sum_{B_i - B_i^\prime = \Delta_2} \lambda_i \nonumber \;.
\eeq
Otherwise, the mass dimension of the leading PQ breaking scalar operator is significantly reduced. A similar result has been obtained in~\cite{Cheung:2010hk}. 

Minimal models (in terms of fermion content) would consist of two pairs of fermion exotics in different representations $\lambda_1$ and $\lambda_2$ of SU(2)$_L$. With two scalars, $\Phi_{\lambda_1}$ and $\Phi_{\lambda_2}$, giving Yukawa masses to each pair of fermions, there are two global symmetries involving the exotics: one global U(1)${}_{\rm PQ}$ involving transformations of $\{\Phi_{\lambda_1},\lambda_1{}_L,\lambda_1{}_R\}$  and one global U(1)$'_{\rm PQ}$ involving $\{\Phi_{\lambda_2},\lambda_2{}_L,\lambda_2{}_R\}$. The leading PQ violating scalar operator will be at most mass dimension $(\lambda_1+\lambda_2)$ and for suitably large SU(2)${}_L$ representations one can forbid PQ violating operators to arbitrary order. 
To forbid all operators up to dimension 10, one requires $\lambda_i \geq 6$ for at least one of the SU(2)${}_L$ representations (since $\lambda_1 = \lambda_2 = 5$ cannot cancel the anomalies) and suitable charge assignments exist to give anomaly-free spectra.
A potential concern however is that large SU(2)${}_L$ representations may lead to Landau poles below the Planck scale and requiring $\alpha_i(M_{\rm Pl})<1$ restricts $\lambda_i\leq 3$ for $f_a\sim10^{11}$~GeV and fermion pairs transforming in the fundamental representation of SU(3), compare the setup in Eq.~\eqref{eq:generalSolution}.\footnote{To be more specific, up to two pairs of SU(2)${}_L$ triplets are viable, or one pair of SU(2)${}_L$ triplets with additional SU(2)${}_L$ doublets.}

From our general discussion above it is clear that the dimension of the leading PQ violating scalar operator can typically be increased by introducing further exotic fermions. Models with fermions up to the adjoint representation of SU(2)${}_L$ necessarily require several exotic fermions to provide adequate protection. Below we outline a viable model using four pairs of fermion exotics in which U(1)${}_{\mathscr{B}}$ gauge invariance protects the PQ symmetry from Planck-scale operators,
 \beq
T^1_L &=(\mathbf{3},\mathbf{3},1,\nicefrac{1}{2}), &T^1_R &=(\mathbf{3},\mathbf{3},1,\nicefrac{5}{8}),\\ 
T^2_L &=(\mathbf{3},\mathbf{3},1,\nicefrac{1}{4}), &T^2_R &=(\mathbf{3},\mathbf{3},1,\nicefrac{3}{8}), \\
S^1_L &=(\mathbf{3},\mathbf{1},1,0)_{\times 3}, &S^1_R &=(\mathbf{3},\mathbf{1},1,\nicefrac{1}{8})_{\times 3}, \\
S^2_L &=(\mathbf{3},\mathbf{1},-\nicefrac{5}{3},\nicefrac{3}{8}), &S^2_R &=(\mathbf{3},\mathbf{1},-\nicefrac{5}{3},-\nicefrac{3}{4}).
\eeq
Yukawa-like mass terms can be constructed for all of the fermion exotics with two Standard Model singlet scalar fields  $\Phi_1$ and  $\Phi_2$ with gauge quantum numbers
\beq
\Phi_1=(\mathbf{1},\mathbf{1},0,-\nicefrac{1}{8}),
\hspace{4mm}\Phi_2=(\mathbf{1},\mathbf{1},0,\nicefrac{9}{8}).
\eeq
 The Yukawa terms of the exotic fields have the form
\beq 
\hspace{-4mm}
\mathcal{L}_{\sharp1} \supset&~
 \Phi_1\left(\kappa_{1} \overline{T^1_L} T^1_R
 +\kappa_{2} \overline{T^2_L} T^2_R+\kappa_{3} \overline{S^1_L} S^1_R\right)\\
&\hspace{10mm} + \kappa_{4}  \Phi_2 \overline{S^2_L} S^2_R + \kappa_{\times}\Phi_1^\dagger \overline{T^1_L} T^2_R+{\rm h.c.} 
\eeq
There are two global symmetries of $\mathcal{L}_{\sharp1}$ involving the exotics, one global U(1)${}_{\rm PQ}$ involves transformations of the fields $\{T^i_L,T^i_R,S^1_L, S^1_R,\Phi_1\}$. These exotic fermions are required to transform together since they all interact with the same scalar. The second U(1)$'_{\rm PQ}$ involves the other exotics $\{S^2_L,S^2_R,\Phi_2\}$.  Both U(1)'s exhibit QCD anomalies,  appropriate for implementing the PQ mechanism
\beq
\mathcal{A}_{{\rm U(1)}_{\rm PQ}\times{\rm SU(3)}^2}=-\mathcal{A}_{{\rm U(1)}'_{\rm PQ}\times{\rm SU(3)}^2} =-\nicefrac{3}{16}\,.
\eeq
The VEVs of $\Phi_1$ and $\Phi_2$ spontaneously break U(1)${}_{\mathscr{B}}$ and both the global PQ symmetries while global baryon number U(1)${}_B$ remains unbroken up to Planck-scale violation and thus typically there is no concern regarding observable proton decay.

For Model $\sharp1$, the leading PQ violating operator is at mass dimension 10,  given by
\beq
\mathcal{O}_{\sharp1}\sim&\frac{1}{9!}\frac{g_{10}}{M_{\rm Pl}^{6}} \Phi_1^9 \Phi_2\,.
\label{bop}
\eeq
Planck-scale breaking of the PQ symmetry due to the operator of eq.~(\ref{bop}) leads to non-zero $\bar\theta$. 
Since the scalar PQ breaking operators are forbidden until mass dimension 10, the value of $\bar\theta$ can naturally be below its experimental limit without excessive tuning of the coupling $g_{10}$ for $f_a \sim \unit[10^{11}]{GeV}$.
Nevertheless these models typically imply a non-zero $\bar\theta$ near the experimental bound. Given that the current experimental limit will improve in the next decade to $d_n\sim 10^{-28}e$~cm~\citep{Hewett:2012ns}, it is quite conceivable that a non-zero neutron electric dipole moment may be observed in the near future.

If some fermion exotics obtain masses other than from Yukawa terms, then one can suppress Planck-scale PQ violation with less fermion exotics. 
For example, an anomaly-free set of exotics is
\beq
X^1_L&=(\mathbf{3},\mathbf{3},\nicefrac{1}{3},-\nicefrac{1}{2}), &X^1_R &=(\mathbf{3},\mathbf{3},\nicefrac{1}{3},\nicefrac{1}{2}), \\[5pt]
X^2_L&=(\mathbf{3},\mathbf{3},\nicefrac{1}{3},\nicefrac{1}{6}), &X^2_R &=(\mathbf{3},\mathbf{3},\nicefrac{1}{3},-\nicefrac{7}{12}), \\[5pt] 
Y_L&=(\mathbf{3},\mathbf{1},\nicefrac{5}{3},\nicefrac{1}{2}), &Y_R &=(\mathbf{3},\mathbf{1},\nicefrac{5}{3},-\nicefrac{1}{4}) \,.
\eeq
One can give (TeV scale) masses to the $X^2_{L,R}$ and $Y_{L,R}$ via higher-dimensional operators through the Lagrangian
\beq
\mathcal{L}_{\sharp2} \supset \kappa_1  \varphi_1 \overline{X^1_L} X^1_R + \frac{\kappa_2}{M_{\rm Pl}}  \varphi_2^2 \overline{X^2_L} X^2_R + \frac{\kappa_3}{M_{\rm Pl}}  \varphi_2^2 \overline{Y_L} Y_R
+ \text{h.c.} \;,
\eeq
where the two scalars have the following charges
\begin{equation}
\varphi_1=(\mathbf{1},\mathbf{1},0,-1), 
\hspace{6mm}
\varphi_2=(\mathbf{1},\mathbf{1},0,\nicefrac{3}{8}).
\end{equation}
The leading PQ violating operator with these scalars is $\varphi_1^3 \varphi_2^8$ at dimension 11.

Furthermore, simple variants can also provide protection to similar order or higher. Models for example with one SU(2)${}_L$ triplet pair and 4 pairs of SU(2)${}_L$ doublets avoids Landau poles below the Planck scale for $f_a = \unit[10^{11}]{GeV}$ and can lead to scenarios in which the leading PQ violating operators only occur at mass dimension 11 or higher. 

\vspace{-1.5ex}

\section{Discussion}
\label{sec:Discussion}

\vspace{-1ex}

Let us briefly discuss the cosmological and phenomenological implications of these models. Axion cosmology significantly depends on whether the PQ symmetry is broken before or after an inflationary phase. In the first case, axions are mainly produced (non-relativistically) via the misalignment mechanism and the axion abundance depends on the initial misalignment angle, allowing for a large range in $f_a$. In the latter case, there is no such freedom and the PQ scale is fixed to about $f_a \sim 10^{11}$~GeV if the axions are required to comprise all of the dark matter~\citep{Borsanyi:2016ksw}.

After PQ breaking a discrete symmetry will be left unbroken, leading to cosmological domain walls which are potentially problematic as they evolve to dominate the energy density of the Universe~\citep{Sikivie:1982qv}.  If PQ breaking occurs prior to inflation the issue is resolved, as the walls are absent after reheating. PQ violating Planck-scale operators with dimension  $D\lesssim12$ offer an alternative solution since they explicitly break the discrete symmetry, lifting the vacuum degeneracy~\citep{Barr:1992qq}, and thus allow for scenarios of post-inflationary PQ breaking. The viability of such scenarios has been recently reanalysed in \cite{Kawasaki:2014sqa,Ringwald:2015dsf}.

Furthermore, the lightest exotics will typically be cosmologically stable, since their quantum numbers forbid low dimension operators via which they can decay. For exotics with masses of order of the PQ scale $f_a$ with a thermal abundance in the early universe, they will contribute a sizable component to the matter abundance. The abundance can be drastically reduced if the reheat temperature after inflation is well below the mass of the exotics such that they are never produced, or if their abundance is diluted via an entropy injection after freeze-out, see e.g.~\citep{Bramante:2017obj, Hamdan:2017psw}.  

Moreover, for exotics whose masses are in the TeV range (as is naturally the case if it is induced by a dimension 5 operator as in our example above) there is the interesting possibility that they may freeze out to a very small abundance due to the large annihilation cross section. However, since the exotics are coloured and typically carry electric charge their abundance is strongly constrained by searches, for instance through the study of anomalous heavy isotopes \citep{Burdin:2014xma}, and even small abundances may be inconsistent with data (for a recent discussion see \cite{DiLuzio:2016sbl,DiLuzio:2017pfr}).

Other constraints from direct searches, astrophysics, and cosmology will be largely similar to other KSVZ-type models~\citep{Patrignani:2016xqp}, although since the exotics are in non-trivial SU(2)${}_L$ representations there may be interesting deviations. In future work we will examine further aspects of model building and phenomenology.

In summary, we have highlighted that U(1)${}_{\rm PQ}$ can arise as a consequence of gauged baryon number U(1)${}_{\mathscr{B}}$. Exotic fermions are required to cancel the gauge anomalies of local baryon number and we have shown that they naturally implement the KSVZ mechanism. Thus these models elegantly intertwine the KSVZ sector with a new baryonic gauge symmetry. In particular the U(1)${}_{\mathscr{B}}$ gauge invariance can protect the PQ symmetry from problematic Planck-scale symmetry breaking which otherwise spoil the PQ solution of the Strong CP Problem.

\subsection*{Acknowledgements}

\vspace{-1ex}

This work is supported by the German Science Foundation (DFG) under the Collaborative Research Center (SFB) 676 Particles, Strings and the Early Universe as well as the ERC Starting Grant `NewAve' (638528).  JU is partially supported by the Alexander von Humboldt Foundation.
  We thank Bogdan Dobrescu, Mads Frandsen, and Andreas Ringwald for helpful conversations.



\begin{thebibliography}{}
  
  \bibitem[Crewther et al.(1979)]{Crewther:1979pi}
  R.~J.~Crewther, P.~Di Vecchia, G.~Veneziano and E.~Witten,
  Phys.\ Lett.\  {\bf 88B} (1979) 123
  [Erratum: Phys.\ Lett.\  {\bf 91B} (1980) 487].
  
  \bibitem[Pendlebury et al.(2015)]{Afach:2015sja}
  J.~M.~Pendlebury {\it et al.},
  Phys.\ Rev.\ D {\bf 92} (2015) 092003
  [\href{https://arxiv.org/abs/1509.04411}{1509.04411}].
 
  \bibitem[Peccei \& Quinn(1977)]{Peccei:1977hh}
  R.~D.~Peccei and H.~R.~Quinn,
  Phys.\ Rev.\ Lett.\  {\bf 38} (1977) 1440.
  %
  \bibitem[Peccei \& Quinn(1977)]{Peccei:1977ur}
  R.~D.~Peccei and H.~R.~Quinn,
  Phys.\ Rev.\ D {\bf 16} (1977) 1791.
  
  \bibitem[Banks \& Seiberg(2010)]{Banks:2010zn}
  T.~Banks and N.~Seiberg,
  Phys.\ Rev.\ D {\bf 83} (2011) 084019
  [\href{https://arxiv.org/abs/1011.5120}{1011.5120}].
  
  \bibitem[Kamionkowski \& March-Russell(1992)]{Kamionkowski:1992mf}
  M.~Kamionkowski and J.~March-Russell,
  Phys.\ Lett.\ B {\bf 282} (1992) 137
  [\href{https://arxiv.org/abs/hep-th/9202003}{hep-th/9202003}].

  \bibitem[Holman et al.(1992)]{Holman:1992us}
  R.~Holman \textit{ et al.}, 
  Phys.\ Lett.\ B {\bf 282} (1992) 132
  [\href{https://arxiv.org/abs/hep-ph/9203206}{hep-ph/9203206}].
  
  \bibitem[Georgi-Hall-Wise(1981)]{Georgi:1981pu}
  H.~M.~Georgi, L.~J.~Hall and M.~B.~Wise,
  Nucl.\ Phys.\ B {\bf 192} (1981) 409.
 
  \bibitem[Randall(1992)]{Randall:1992ut}
  L.~Randall,
  Phys.\ Lett.\ B {\bf 284} (1992) 77.

  \bibitem[Barr \& Seckel(1992)]{Barr:1992qq}
  S.~M.~Barr and D.~Seckel,
  Phys.\ Rev.\ D {\bf 46} (1992) 539.

  \bibitem[Redi \& Sato(2016)]{Redi:2016esr}
  M.~Redi and R.~Sato,
  JHEP {\bf 05} (2016) 104
  [\href{https://arxiv.org/abs/1602.05427}{1602.05427}].
  
  \bibitem[Di Luzio-Nardi-Ubaldi(2017)]{DiLuzio:2017tjx}
  L.~Di Luzio, E.~Nardi and L.~Ubaldi,
  Phys.\ Rev.\ Lett.\  {\bf 119} (2017) 011801
  [\href{https://arxiv.org/abs/1704.01122}{1704.01122}].
  
  \bibitem[Lee \& Yang(1955)] {Lee:1955vk}
  T.~D.~Lee and C.~N.~Yang,
  Phys.\ Rev.\  {\bf 98} (1955) 1501.
   
  \bibitem[Pais(1973)]{Pais:1973mi}
  A.~Pais,
  Phys.\ Rev.\ D {\bf 8} (1973) 1844.
  
  \bibitem[Foot-Joshi-Lew(1989)]{Foot:1989ts}
  R.~Foot, G.~C.~Joshi and H.~Lew,
  Phys.\ Rev.\ D {\bf 40} (1989) 2487.

  \bibitem[Fileviez Perez \& Wise(2010)]{FileviezPerez:2010gw}
  P.~Fileviez Perez and M.~B.~Wise,
  Phys.\ Rev.\ D {\bf 82} (2010) 011901
  [Erratum: Phys.\ Rev.\ D {\bf 82} (2010) 079901]
  [\href{https://arxiv.org/abs/1002.1754}{1002.1754}].
  
  \bibitem[Fileviez Perez \& Wise(2011)]{FileviezPerez:2011pt}
  P.~Fileviez Perez and M.~B.~Wise,
  JHEP {\bf 08} (2011) 068
  [\href{https://arxiv.org/abs/1106.0343}{1106.0343}].
  
  \bibitem[Duerr-Fileviez Perez-Wise(2013)]{Duerr:2013dza}
  M.~Duerr, P.~Fileviez Perez and M.~B.~Wise,
  Phys.\ Rev.\ Lett.\  {\bf 110} (2013) 231801
  [\href{https://arxiv.org/abs/1304.0576}{1304.0576}].

  \bibitem[Kim(1979)]{Kim:1979if}
  J.~E.~Kim,
  Phys.\ Rev.\ Lett.\  {\bf 43} (1979) 103.
  
  \bibitem[Shifman-Vainshtein-Zakharov(1979)]{Shifman:1979if}
  M.~A.~Shifman, A.~I.~Vainshtein and V.~I.~Zakharov,
  Nucl.\ Phys.\ B {\bf 166} (1980) 493.
  
  \bibitem[Fukuda et al.(2017)]{Fukuda:2017ylt}
  H.~Fukuda, M.~Ibe, M.~Suzuki and T.~T.~Yanagida,
  Phys.\ Lett.\ B {\bf 771} (2017) 327
  [\href{https://arxiv.org/abs/1703.01112}{1703.01112}].
  
  \bibitem[Ringwald et al.(2016)]{Patrignani:2016xqp}
  A.~Ringwald, L.~J.~Rosenberg, and G.~Rybka, \textit{in} 
  C.~Patrignani {\it et al.} [{\tt Particle Data Group}],
  Chin.\ Phys.\ C {\bf 40} (2016)  100001.
  
  \bibitem[Kallosh et al.(1995)]{Kallosh:1995hi}
  R.~Kallosh, A.~D.~Linde, D.~A.~Linde and L.~Susskind,
  Phys.\ Rev.\ D {\bf 52} (1995) 912
  [\href{https://arxiv.org/abs/hep-th/9502069}{hep-th/9502069}].

  \bibitem[Dobrescu(1996)]{Dobrescu:1996jp}
  B.~A.~Dobrescu,
  Phys.\ Rev.\ D {\bf 55} (1997) 5826
  [\href{https://arxiv.org/abs/hep-ph/9609221}{hep-ph/9609221}].

  \bibitem[Cheung(2010)]{Cheung:2010hk}
  C.~Cheung,
  JHEP {\bf 06} (2010) 074
  [\href{https://arxiv.org/abs/1003.0941}{1003.0941}].

  \bibitem[Hewett et al.(2012)]{Hewett:2012ns}
  J.~L.~Hewett {\it et al.},
  [\href{https://arxiv.org/abs/1205.2671}{1205.2671}].

  \bibitem[Borsanyi et al.(2016)]{Borsanyi:2016ksw}
  S.~Borsanyi {\it et al.},
  Nature {\bf 539} (2016)  69
  [\href{https://arxiv.org/abs/1606.07494}{1606.07494}].

  \bibitem[Sikivie(1982)]{Sikivie:1982qv}
  P.~Sikivie,
  Phys.\ Rev.\ Lett.\  {\bf 48} (1982) 1156.

  \bibitem{Kawasaki:2014sqa}
  M.~Kawasaki, K.~Saikawa and T.~Sekiguchi,
  Phys.\ Rev.\ D {\bf 91} (2015) 065014
  [\href{https://arxiv.org/abs/1412.0789}{1412.0789}].
  
  \bibitem{Ringwald:2015dsf}
  A.~Ringwald and K.~Saikawa,
  Phys.\ Rev.\ D {\bf 93} (2016) 085031
  [Addendum: Phys.\ Rev.\ D {\bf 94} (2016) 049908]
  [\href{https://arxiv.org/abs/1512.06436}{1512.06436}].
  
  \bibitem[Bramante \& Unwin(2017)]{Bramante:2017obj}
  J.~Bramante and J.~Unwin,
  JHEP {\bf 02} (2017) 119
  [\href{https://arxiv.org/abs/1701.05859}{1701.05859}].

  \bibitem[Hamdan \& Unwin(2017)]{Hamdan:2017psw}
  S.~Hamdan and J.~Unwin,
  [\href{https://arxiv.org/abs/1710.03758}{1710.03758}].
  
  \bibitem[Burdin et al.(2014)]{Burdin:2014xma}
  S.~Burdin \textit{et al.}
  Phys.\ Rept.\  {\bf 582} (2015) 1
  [\href{https://arxiv.org/abs/1410.1374}{1410.1374}].
  
  \bibitem[Di Luzio-Mescia-Nardi(2016)]{DiLuzio:2016sbl}
  L.~Di Luzio, F.~Mescia and E.~Nardi,
  Phys.\ Rev.\ Lett.\  {\bf 118} (2017) 031801
  [\href{https://arxiv.org/abs/1610.07593}{1610.07593}].
  
  \bibitem[Di Luzio-Mescia-Nardi(2017)]{DiLuzio:2017pfr}
  L.~Di Luzio, F.~Mescia and E.~Nardi,
  Phys.\ Rev.\ D {\bf 96} (2017) 075003
  [\href{https://arxiv.org/abs/1705.05370}{1705.05370}].
\end{thebibliography}
\end{document}